\documentstyle[11pt]{article}
\begin{document}
\title{\bf DIFFERENTIAL INVARIANTS\\ FOR HIGHER--RANK TENSORS.\\ A PROGRESS REPORT}

\author{{\bf Victor Tapia}\footnote{{\tt tapiens@hotmail.com}}\\
\\
{\it Departamento de Matem\'aticas}\\
{\it Universidad Nacional de Colombia}\\
{\it Bogot\'a, Colombia}\\}

\maketitle

\begin{abstract}

We outline the construction of differential invariants for higher--rank tensors.

\end{abstract}

\section{ Introduction}

Higher spin fields might help in a unified description of physical interactions. Higher--spin fields were introduced in \cite{FF,Fro}, and has been considered since then in several contexts \cite{BB,BBV,Cur,DD1,DD2,DF,FS,Hul,MH1,MH2,Vas}. One possibility to describe higher--spin fields is by means of higher--rank tensors, which has been also considered in alternative gravitational theories \cite{Ta1,TRMC,TR}. The first step in any field theoretical description of this kind is the construction of a geometrical invariant to be used as a Lagrangian. The main line of attack has been to consider higher--rank tensors in a Minkowski or Riemannian background. On the other hand, it is interesting to consider field theories constructed from the higher--rank tensors alone, that is, to develope the ``differential geometry'' associated to higher--rank tensors in a way similar to the way in which Riemannian geometry is constructed from a second--rank tensor (a metric) \cite{DD2}. Therefore, we need to construct differential invariants for higher--rank tensors. There are several general results concerning the construction of differential invariants for tensors \cite{ABH,FKWC,MV,Tho,WM}. The first step in the construction of differential invariants is to determine the number of functionally independent invariants which can be constructed out from a given tensor and its derivatives. We restrict our considerations to completely symmetric higher--rank tensors. In that case, the simplest differential invariant which can be constructed contains derivatives of an order equal to the rank of the tensor. The second step is the explicit construction of these differential invariants. For first--rank tensors (vectors) the solution is the Maxwell tensor while for second--rank tensors (metrics) the solution is the Riemann--Christoffel tensor. However, for higher--rank tensors the method faces several practical obstructions due, mainly, to the fact that the inverse higher--rank tensor is an involved algebraic function of the original tensor \cite{Ta3}. Therefore, the few existing considerations have been restricted to linearised quantities \cite{DD1}; see also
\cite{Hul,MH1,MH2}.

Therefore, the construction of differential invariants for higher--rank tensors is still an open problem.

The work is organised as follows. In section 2 we outline the general method for the construction of differential invariants. A first result is that the simplest tensor differential invariant contains derivatives of the same order as the rank of the tensor. In section 3 we review the construction for first--rank tensors (vectors) and second--rank tensors (metrics). In section 4 we outline the same construction for higher--rank tensors.

\section{The Number of Differential Invariants\\ for Completely Symmetric Tensors}

In this work we adopt the taxonomic definition of a tensor based on the transformation rule of its components: {\it a tensor is something which transforms like a tensor}. An $r$th--rank covariant tensor ${\bf G}$ is an object such that its components $G_{i_1\cdots i_r}$ transform like

\begin{equation}
G_{a_1\cdots a_r}({\bf y})={X^{i_1}}_{a_1}\,\cdots\,{X^{i_r}}_{a_r}\,G_{i_1\cdots i_r}({\bf x})\,,
\label{1}
\end{equation}

\noindent where

\begin{equation}
{X^i}_a={{\partial x^i}\over{\partial y^a}}\,.
\label{2}
\end{equation}

\noindent For later convenience let us also introduce

\begin{equation}
{X^i}_{ab}={{\partial^2x^i}\over{\partial y^a\partial y^b}}\,,
\label{3}
\end{equation}

\noindent with obvious extensions to higher order derivatives, and

\begin{equation}
{Y^a}_i={{\partial y^a}\over{\partial x^i}}\,.
\label{4}
\end{equation}

\noindent Therefore, a tensor is an object such that in the transformation rule of its components only the transformation matrix ${X^i}_a$ appears. When considering derivatives of a tensor, derivatives of the transformation matrix will appear in the corresponding transformation rules; these illegal terms show that in general the derivative of a tensor is not a tensor. However, by means of only symmetrisation operations we can construct combination of derivatives not containing illegal terms. In order to determine the number of relations of this kind we must count the relations and the illegal terms.

The derivative of (\ref{1}) is given by

\begin{eqnarray}
\partial_c G_{a_1\cdots a_r}({\bf y})&=&{X^{i_1}}_{a_1}\,\cdots\,{X^{i_r}}_{a_r}\,{X^j}_c\,\partial_j G_{i_1\cdots i_r}({\bf x})\nonumber\\
&&+\left({X^{i_1}}_{a_1c}\,\cdots\,{X^{i_r}}_{a_r}+\cdots+{X^{i_1}}_{a_1}\,\cdots\,{X^{i_r}}_{a_r c}
\right)\,G_{i_1\cdots i_r}({\bf x})\,.
\label{5}
\end{eqnarray}

\noindent The number of relations $E(1,n,r)$ in (\ref{5}) is the number of derivatives, $n$, times the number of components $T(n,r)$ of ${\bf G}$, given by

\begin{equation}
T(n,\,r)={{(n+r-1)!}\over{(n-1)!r!}}\,.
\label{6}
\end{equation}

\noindent Therefore

\begin{equation}
E(1,\,n,\,r)=n\,\cdot\,T(n,\,r)\,.
\label{7}
\end{equation}

\noindent The illegal terms in (\ref{5}) are $(\partial^2X)$ given by

\begin{equation}
(\partial^2X)_{c a_1\cdots a_r}={X^{i_1}}_{a_1c}\,\cdots\,{X^{i_r}}_{a_r}\,G_{i_1\cdots i_r}=
{X^i}_{a_1c}\,{Y^b}_i\,G_{b a_2\cdots a_r}\,.
\label{8}
\end{equation}

\noindent The number $U(1,n,r)$ of illegal terms is given by the number of symmetrised derivatives on $X$, that is $n(n+1)/2$, times the symmetries over $r-1$ indices in ${\bf G}$, that is $T(n,r-1)$. Then,

\begin{equation}
U(1,\,n,\,r)={{n(n+1)}\over2}\,\cdot\,T(n,\,r-1)\,.
\label{9}
\end{equation}

\noindent Even when the illegal terms are ${X^i}_{ab}$ they appear always in the combination shown in (\ref{8}) and therefore the counting of illegal terms is as shown above. For a further derivative we obtain

\begin{equation}
\partial_{c_1c_2}G_{a_1\cdots a_r}({\bf y})=\left[{X^{i_1}}_{a_1c_1c_2}\,\cdots\,{X^{i_r}}_{a_r}+\cdots+{X^{i_1}}_{a_1}\,
\cdots\,{X^{i_r}}_{a_rc_1c_2}\right]\,G_{i_1\cdots i_r}({\bf x})+\cdots\,.
\label{10}
\end{equation}

\noindent The counting of relations and illegal terms, $(\partial^3X)$, in (\ref{10}) is as above and they are given by

\begin{eqnarray}
E(2,\,n,\,r)&=&{{n(n+1)}\over2}\,\cdot\,T(n,\,r)\,,
\label{11}\\
U(2,\,n,\,r)&=&{{n(n+1)(n+2)}\over{3!}}\,\cdot\,T(n,\,r-1)\,.
\label{12}
\end{eqnarray}

\noindent When considering $d$th--order derivatives the number of relations and the number of illegal terms, $(\partial^{d+1}X)$, are given by

\begin{eqnarray}
E(d,\,n,\,r)&=&{{(n+d-1)!}\over{(n-1)!d!}}\,\cdot\,T(n,\,r)={{(n+d-1)!}\over{(n-1)!d!}}\,
\cdot\,{{(n+r-1)!}\over{(n-1)!r!}}\,,
\label{13}\\
U(d,\,n,\,r)&=&{{(n+d)!}\over{(n-1)!(d+1)!}}\,\cdot\,T(n,\,r-1)=
{{(n+d)!}\over{(n-1)!(d+1)!}}\,\cdot\,{{(n+r-2)!}\over{(n-1)!(r-1)!}}\,.
\label{14}
\end{eqnarray}

The possibility of finding differential invariants depends on the relative values of $E(d,n,r)$ and $U(d,n,r)$. The difference of these two quantities is

\begin{equation}
\Delta(d,\,n,\,r)=E(d,\,n,\,r)-U(d,\,n,\,r)={{(n+d-1)!(n+r-2)!}\over{[(n-1)!]^2(d+1)!r!}}\,(n-1)\,(d-r+1)\,.
\label{15}
\end{equation}

\noindent If $d<r-1$, then $\Delta<0$, and there are more illegal terms than relations. Therefore, it is not possible to find relations not involving the illegal terms. If $d=r$, $\Delta=0$, and the number of relations and the number of illegal terms are equal; then it is possible to solve for the illegal terms but still there is no relation not involving them. When considering a further derivative, $d=r$, $\Delta>0$, we have more relations than illegal terms. Therefore we obtain a differential invariant ${\bf R}$ with a number of components given by

\begin{eqnarray}
R(n,\,r)&=&E(r,\,n,\,r)-U(r,\,n,\,r)\nonumber\\
&=&\left({{(n+r-1)!}\over{r!(n-1)!}}\right)^2-{{(n+r)!}\over{(n-1)!(r+1)!}}\,\cdot\,{{(n+r-2)!}\over{(n-1)!(r-1)!}}
\nonumber\\
&=&{{(n+r-1)!(n+r-2)!}\over{(n-1)!(n-2)!(r+1)!r!}}\,.
\label{16}
\end{eqnarray}

\noindent Let us consider the above expression for special values of $r$ and $n$. For the first values of $r$ we obtain

\begin{eqnarray}
R(n,\,1)&=&{1\over2}\,n\,(n-1)\,,
\label{17}\\
R(n,\,2)&=&{1\over{12}}\,n^2\,(n^2-1)\,,
\label{18}\\
R(n,\,3)&=&{1\over{144}}\,n^2\,(n+1)^2\,(n^2+n-2)\,,
\label{181}\\
R(n,\,4)&=&{1\over{2880}}\,n^2\,(n+1)^2\,(n+2)^2\,(n^2+2\,n-3)\,,
\label{182}
\end{eqnarray}

\noindent and the resulting differential invariants are the Maxwell tensor for $r=1$ and the Riemann--Christoffel tensor for $r=2$. However, for the first values of $n$ we obtain simpler expressions.

\begin{eqnarray}
R(1,\,r)&=&0\,,
\label{183}\\
R(2,\,r)&=&1\,,
\label{184}\\
R(3,\,r)&=&{1\over2}\,(r+2)\,[(r+2)-1]\,,
\label{185}\\
R(4,\,r)&=&{1\over{12}}\,(r+2)^2\,[(r+2)^2-1]\,.
\label{186}
\end{eqnarray}

The expression (\ref{16}) can be rewritten as

\begin{equation}
R(n,\,r)={{(N+r-1)!}\over{(N-1)!r!}}-{{(N+r-3)!}\over{(N-1)!(r-2)!}}\,\cdot\,{{n!}\over{(n-4)!4!}}\,,
\label{19}
\end{equation}

\noindent where $N=n(n-1)/2$, which is the number of components of a $(2r)$th--rank tensor ${\bf R}$ with components $R_{i_1j_1\cdots i_r j_r}$ with the following symmetries: Indices are ordered in antisymmetric couples, that is

\begin{equation}
R_{i_1j_1\cdots i_r j_r}=R_{[i_1j_1]\cdots[i_r j_r]}\,.
\label{20}
\end{equation}

\noindent Furthermore, the tensor is completely symmetric with respect to the couples of indices $[ij]$, which is the first term in (\ref{19}). The second term means that cyclic permutation over 3 indices in two couples, $n!/(n-4)!4!$, is zero; the first part of the second term, $(N+r-3)!/(N-1)!(r-2)!$, is the number of ways in which this choice can be made.

We must still consider the case $d>r$, $\Delta>0$. In this case the corresponding differential invariants can be expressed in terms of derivatives of ${\bf R}$. It is however still interesting to consider the corresponding number of invariants, given by (\ref{15}). In the second--rank case, $r=2$, the corresponding expression reduces to

\begin{equation}
\Delta(d,\,n,\,2)={1\over2}\,n\,(d-1)\,{{(n+d-1)!}\over{(n-2)!(d+1)!}}\,.
\label{21}
\end{equation}

\noindent Scalar invariants are obtained from (\ref{21}) just by substracting the $n(n-1)/2$ conditions which fix a local Lorentz transformation. We obtain

\begin{equation}
S(n,\,d)=R(n,\,d)-{1\over2}\,n\,(n-1)={1\over2}\,n\,(d-1)\,{{(n+d-1)!}\over{(n-2)!(d+1)!}}-{1
\over2}\,n\,(n-1)\,.
\label{22}
\end{equation}

\noindent This result coincides with that in \cite{MV} for scalar invariants. Our formulae (\ref{21}) and (\ref{22}) have been obtained by means of a simple counting of relations and illegal terms and therefore our procedure is clearer than the one used in \cite{MV}. Furthermore, our formula (\ref{22}), even when numerically equivalent, is simpler than that in \cite{MV}.

\section{Explicit Construction of Invariants}

Let us start by considering a vector field ${\bf A}$ with components $A_i$. The corresponding transformation rules are

\begin{equation}
A_a({\bf y})={X^i}_a\,A_i({\bf x})\,.
\label{23}
\end{equation}

\noindent The derivative of (\ref{23}) is given by

\begin{equation}
\partial_b A_a({\bf y})={X^i}_a\,{X^j}_b\,\partial_j A_i({\bf x})+{X^i}_{ab}\,A_i({\bf x})\,.
\label{24}
\end{equation}

\noindent Therefore, the derivative (\ref{24}) is not a tensor. The number of relations in (\ref{24}) is $n^2$, while the number of illegal terms, $(\partial^2X)$, is $n(n+1)/2$. Since (\ref{24}) is a linear algebraic system of equations there must be $n^2-n(n+1)/2=n(n-1)/2$ relations not involving the illegal terms $(\partial^2X)$. In fact, we have

\begin{equation}
F_{ab}({\bf y})={X^i}_a\,{X^j}_b\,F_{ij}({\bf x})\,,
\label{25}
\end{equation}

\noindent where

\begin{equation}
F_{ij}=\partial_i A_j-\partial_j A_i\,.
\label{26}
\end{equation}

\noindent which we recognise as the Maxwell tensor.

\bigskip

Let us now consider the same construction for a second--rank symmetric tensor ${\bf g}$, a mtric, with components $g_{ij}$. The first result for a metric was obtained by Gauss \cite{Gau} for $n=2$. In {\tt 1861} Riemann constructed \cite{Rie} what is today known as the Riemann--Christoffel tensor. Let us start by reminding some simple results. The inverse metric ${\bf g}^{-1}$ is defined as the tensor with components $g^{ij}$ satisfying

\begin{equation}
g^{ik}\,g_{jk}=\delta^i_j\,.
\label{27}
\end{equation}

\noindent This is not only the definition of the inverse metric ${\bf g}^{-1}$ but also a linear algebraic system of equations; we have $n^2$ equations and $n^2$ unknowns, threfore the system has a unique solution. The determinant of the metric is given by

\begin{equation}
g=\det(g_{ij})={1\over{n!}}\,\epsilon^{i_1\cdots i_n}\,\epsilon^{j_1\cdots j_n}\,g_{i_1j_1}\,\cdots\,g_{i_n j_n}\,.
\label{28}
\end{equation}

\noindent The condition for (\ref{27}) to have a solution is $g\not=0$. In this case we can define

\begin{equation}
g^{ij}={1\over g}\,{1\over{(n-1)!}}\,\epsilon^{i i_1\cdots i_{n-1}}\,\epsilon^{j j_1\cdots j_{n-1}}\,g_{i_1j_1}\,\cdots\,g_{i_{n-1}j_{n-1}}\,,
\label{29}
\end{equation}

\noindent which satisfies (\ref{27}). Therefore, $g^{ij}$, defined as in (\ref{29}), is the inverse metric.

The transformation rules for ${\bf g}$ and its derivatives are given by

\begin{eqnarray}
g_{ab}({\bf y})&=&{X^i}_a\,{X^j}_b\,g_{ij}({\bf x})\,,
\label{30}\\
\partial_c g_{ab}({\bf y})&=&\left({X^i}_{ac}\,{X^j}_b+{X^i}_a\,{X^j}_{bc}\right)\,g_{ij}({\bf x})+
{X^i}_a\,{X^j}_b\,{X^k}_c\,\partial_k g_{ij}({\bf x})\,,
\label{31}\\
\partial_{dc}g_{ab}({\bf y})&=&\left[{X^i}_{dca}\,{X^j}_b+{X^i}_a\,{X^j}_{dcb}+{X^i}_{ca}\,{X^j}_{db}
+{X^i}_{da}\,{X^j}_{cb}\right]\,g_{ij}({\bf x})\nonumber\\
&&+\left[\left({X^i}_{ca}\,{X^j}_b+{X^i}_a\,{X^j}_{cb}\right)\,{X^k}_d+\left({X^i}_{da}\,{X^j}_b
+{X^i}_a\,{X^j}_{db}\right)\,{X^k}_c\right.\nonumber\\
&&\left.+{X^i}_a\,{X^j}_b\,{X^k}_{dc}\right]\,\partial_k g_{ij}({\bf x})\nonumber\\
&&+{X^i}_a\,{X^j}_b\,{X^k}_c\,{X^l}_d\,\partial_{lk}g_{ij}({\bf x})\,.
\label{32}
\end{eqnarray}

There are several ways of constructing the invariant ${\bf R}$ through different tensor manipulations. The simplest way is to solve for the terms $(\partial^2X)$; from (\ref{31}) we obtain

\begin{equation}
{X^k}_{ab}={X^k}_c\,{\Gamma^c}_{ab}({\bf y})-{X^i}_a\,{X^j}_b\,{\Gamma^k}_{ij}({\bf x})\,,
\label{33}
\end{equation}

\noindent where

\begin{equation}
{\Gamma^c}_{ab}={1\over2}\,g^{cd}\,\left(\partial_a g_{bd}+\partial_b g_{ad}-\partial_d g_{ab}\right)\,.
\label{34}
\end{equation}

\noindent This manipulation can be done because of the existence of an inverse metric ${\bf g}^{-1}$. Let us now consider a further derivative of (\ref{33}). We obtain

\begin{eqnarray}
{X^k}_{abc}&=&{X^k}_{cd}\,{\Gamma^c}_{ab}({\bf y})+{X^k}_c\,\partial_d{\Gamma^c}_{ab}({\bf y})\nonumber\\
&&-{X^i}_{ac}\,{X^j}_b\,{\Gamma^k}_{ij}({\bf x})-{X^i}_a\,{X^j}_{bc}\,{\Gamma^k}_{ij}({\bf x})-{X^i}_a\,{X^j}_b\,{X^l}_c\,\partial_l{\Gamma^k}_{ij}({\bf x})\,.
\label{35}
\end{eqnarray}

\noindent Using (\ref{33}) we can elliminate the terms $(\partial^2X)$. On the other hand, the left--hand side of (\ref{35}) is completely symmetric and this fact allows to elliminate the derivatives $(\partial^3X)$. We arrive then to the Riemann--Christoffel tensor.

It is obvious that already for this example there is an unavoidable (and, of course, undesired) proliferation of indices. In order to reduce the overabundance of indices and terms in several equations we make some simplifying assumptions.
Let us observe that the invariant we want to construct is of the form

\begin{equation}
{\bf R}=\partial^2{\bf g}+{\bf g}^{-1}\,(\partial{\bf g})^2\,.
\label{36}
\end{equation}

\noindent A sufficient condition for the vanishing of this invariant is that ${\bf g}$ be a constant tensor. Since ${\bf R}$ is a tensor it will also vanishes in a second system of coordinates in which the tensor ${\bf g}$ is no more a constant tensor. In this second system of coordinates ${\bf R}$ is the simplest vanishing differential invariant which can be constructed in this way. Let us therefore choose the metric in the first system of coordinates as constant

\begin{equation}
g_{b_1b_2}={X^i}_{b_1}\,{X^j}_{b_2}\,\eta_{ij}\,.
\label{37}
\end{equation}

\noindent From a practical point of view this choice means that we must not mind about several terms involving derivatives at the other side of the relations. The first derivatives of this expression are given by

\begin{eqnarray}
\partial_c g_{ab}({\bf y})&=&\left[{X^i}_{ac}\,{X^j}_b+{X^i}_a\,{X^j}_{bc}\right]\,\eta_{ij}\,,
\label{38}\\
\partial_{dc}g_{ab}({\bf y})&=&\left[{X^i}_{dca}\,{X^j}_b+{X^i}_a\,{X^j}_{dcb}+{X^i}_{ca}\,{X^j}_{db}
+{X^i}_{da}\,{X^j}_{cb}\right]\,\eta_{ij}\,.
\label{39}
\end{eqnarray}

Of course, all the manipulations above work properly for a second--rank tensor (they are specific). However, what we need is a construction method which can be used also for higher--rank tensors. In order to construct the invariant ${\bf R}$ in a systematic way which will be useful for generalizations to higher--ranks let us remind that the number of components means certain symmetries. Let us therefore consider

\begin{eqnarray}
\partial_{[a_2[a_1}g_{b_1]b_2]}&=&\partial_{a_2a_1}g_{b_1b_2}-\partial_{a_2b_1}g_{a_1b_2}-\partial_{b_2a_1}g_{b_1a_2}
+\partial_{b_2b_1}g_{a_1a_2}\nonumber\\
&=&2\,\left({X^i}_{a_2b_1}\,{X^j}_{a_1b_2}-{X^i}_{a_1a_2}\,{X^j}_{b_1b_2}\right)\,\eta_{ij}\,.
\label{40}
\end{eqnarray}

\noindent Let us now remind that the terms $(\partial^2X)$ can be solved from (\ref{38}). The solution is

\begin{equation}
{X^i}_{b_1b_2}={X^i}_c\,{\Gamma^c}_{b_1b_2}\,.
\label{41}
\end{equation}

\noindent Therefore

\begin{eqnarray}
R_{a_1b_1a_2b_2}&=&{1\over2}\,\left(\partial_{a_2a_1}g_{b_1b_2}-\partial_{a_2b_1}g_{a_1b_2}-\partial_{b_2a_1}g_{b_1a_2}
+\partial_{b_2b_1}g_{a_1a_2}\right)\nonumber\\
&&-g_{cd}\,\left({\Gamma^c}_{a_2b_1}\,{\Gamma^d}_{a_1b_2}-{\Gamma^c}_{a_1a_2}\,{\Gamma^d}_{b_1b_2}\right)=0\,.
\label{42}
\end{eqnarray}

\noindent Therefore, the vanishing of the differential invariant ${\bf R}$ is the integrability condition for ${\bf g}$ to be of the form (\ref{37}).

\bigskip

Now we introduce a simplification in the notation. The indices in the fixed reference frame play no role. Therefore, the expressions above can be simplified as follows. Let us rewrite (\ref{37}) as

\begin{equation}
g_{ab}=X_a\cdot X_b\,.
\label{43}
\end{equation}

\noindent The first derivatives of this expression are given by

\begin{eqnarray}
\partial_c g_{ab}({\bf y})&=&X_{ac}\cdot X_b+X_a\cdot X_{bc}\,,
\label{44}\\
\partial_{dc}g_{ab}({\bf y})&=&X_{dca}\cdot X_b+X_a\cdot X_{dcb}+X_{ca}\cdot X_{db}+X_{da}\cdot X_{cb}\,.
\label{45}
\end{eqnarray}

\noindent The antisymmetric part of (\ref{45}) is given by

\begin{equation}
\partial_{[a_2[a_1}g_{b_1]b_2]}=2\,\left(X_{a_2b_1}\cdot X_{a_1b_2}-X_{a_1a_2}\cdot X_{b_1b_2}\right)\,.
\label{46}
\end{equation}

\noindent which is (\ref{40}) leading to (\ref{42}).

\section{Higher--Rank Tensors}

The possibility of implementing the method exposed previously relies on the possibility of inverting several relations. However, as we will see now, the definition of an inverse tensor for higher--rank tensors is not direct. As shown in \cite{Ta3} only even $r$ can be constructed consistently. In order to fix the ideas, we illustrate them in the fourth--rank case.

Let us consider completely symmetric fourth--rank tensor $G_{ijkl}$. The inverse tensor ${\bf G}^{-1}$ is a tensor with components $G^{ijkl}$ satisfying

\begin{equation}
G^{i k_1k_2k_3}\,G_{j k_1k_2k_3}=\delta^i_j\,.
\label{47}
\end{equation}

\noindent This is the definition of the inverse tensor but now, in contrast with (\ref{27}), there are more unknowns than relations and therefore the solution is not unique. In order to avoid this undeterminacy let us define the inverse tensor in a way similar to (\ref{28}). The determinant of ${\bf G}$ is defined by

\begin{equation}
G=\det(G_{ijkl})={1\over{n!}}\,\epsilon^{i_1\cdots i_n}\,\cdots\,\epsilon^{l_1\cdots l_n}\,G_{i_1j_1k_1l_1}\,\cdots\,G_{i_n j_n k_n l_n}\,.
\label{48}
\end{equation}

\noindent If $G\not=0$ we can define

\begin{equation}
G^{ijkl}={1\over G}\,{1\over{(n-1)!}}\,\epsilon^{i m_1\cdots m_{n-1}}\,\cdots\,\epsilon^{l q_1\cdots q_{n-1}}\,G_{m_1n_1p_1q_1}\,\cdots\,G_{m_{n-1}m_{n-1}p_{n-1}q_{n-1}}\,.
\label{49}
\end{equation}

\noindent This tensor satisfies (\ref{47}). The expression above can be generalized to

\begin{eqnarray}
&&G^{i_1j_1k_1l_1}\,G^{i_2j_2k_2l_2}\nonumber\\
&&-\left(G^{i_2j_1k_1l_1}\,G^{i_1j_2k_2l_2}+G^{i_1j_2k_1l_1}\,G^{i_2j_1k_2l_2}+G^{i_1j_1k_2l_1}\,G^{i_2j_2k_1l_2}+G^{i_1j_1k_1l_2}\,G^{i_2j_2k_2l_1}\right)\nonumber\\
&&+\left(G^{i_2j_2k_1l_1}\,G^{i_1j_1k_2l_2}+G^{i_2j_1k_2l_1}\,G^{i_1j_2k_1l_2}+G^{i_2j_1k_1l_2}\,G^{i_1j_2k_2l_1}\right)\nonumber\\
&=&{1\over G}\,{1\over{(n-2)!}}\,\epsilon^{i_1i_2m_1\cdots m_{n-2}}\,\cdots\,\epsilon^{l_1l_2q_1\cdots q_{n-2}}\,G_{m_1n_1p_1q_1}\,\cdots\,G_{m_{n-2}n_{n-2}p_{n-2}q_{n-2}}\,.
\label{50}
\end{eqnarray}

\noindent Contracting with $G_{i_2j_2k_2l_2}$ we obtain that the inverse tensor (\ref{49}) also satisfies the relation

\begin{eqnarray}
G_{(ij|mn}\,G^{mnpq}\,G_{pq|kl)}=G_{ijkl}\,.
\label{51}
\end{eqnarray}

\noindent By the way, this relation can be used as a better definition for the inverse tensor since now the number of equations and the number of unknowns are equal.

\bigskip

Let us now outline the construction of invariants. Let us start considering the transformation rule for ${\bf G}$, that is,

\begin{equation}
G_{a_1a_2a_3a_4}={X^i}_{a_1}\,{X^j}_{a_2}\,{X^k}_{a_3}\,{X^l}_{a_4}\,G_{ijkl}\,.
\label{52}
\end{equation}

\noindent The invariant we want to construct is of the form

\begin{equation}
{\bf R}=\partial^4{\bf G}+{\bf G}^{-1}\,(\partial^3{\bf G})\,(\partial{\bf G})+{\bf G}^{-1}\,(\partial^2{\bf G})^2+{\bf G}^{-2}\,(\partial^2{\bf G})\,(\partial{\bf G})^2+{\bf G}^{-3}\,(\partial{\bf G})^4\,.
\label{53}
\end{equation}

\noindent A sufficient condition for the vanishing of this invariant is that ${\bf G}$ be a constant tensor. Let us therefore write

\begin{equation}
G_{a_1a_2a_3a_4}={X^i}_{a_1}\,{X^j}_{a_2}\,{X^k}_{a_3}\,{X^l}_{a_4}\,\eta_{ijkl}=X_{a_1}\cdot X_{a_2}\cdot X_{a_3}\cdot X_{a_4}\,,
\label{54}
\end{equation}

\noindent where we assume that $\eta_{ijkl}$ is a constant tensor. Then, as in the second--rank case, the differential invariant we are looking for appears as the integrability condition for ${\bf G}$ to be of the form (\ref{54}). The first derivatives of (\ref{54}) are given by

\begin{eqnarray}
\partial_b G_{a_1a_2a_3a_4}&=&X_{a_1b}\cdot X_{a_2}\cdot X_{a_3}\cdot X_{a_4}+X_{a_1}\cdot X_{a_2b}\cdot X_{a_3}\cdot X_{a_4}\nonumber\\
&&+X_{a_1}\cdot X_{a_2}\cdot X_{a_3b}\cdot X_{a_4}+X_{a_1}\cdot X_{a_2}\cdot X_{a_3}\cdot X_{a_4b}\,,
\label{55}\\
\partial_{b_1b_2}G_{a_1a_2a_3a_4}&=&\left[X_{a_1b_1b_2}\cdot X_{a_2}\cdot X_{a_3}\cdot X_{a_4}+X_{a_1}\cdot X_{a_2b_1b_2}\cdot X_{a_3}\cdot X_{a_4}\right.\nonumber\\
&&+\left.X_{a_1}\cdot X_{a_2}\cdot X_{a_3b_1b_2}\cdot X_{a_4}+X_{a_1}\cdot X_{a_2}\cdot X_{a_3}\cdot X_{a_4b_1b_2}\right]\nonumber\\
&&+\left[X_{a_1(b_1|}\cdot X_{a_2|b_2)}\cdot X_{a_3}\cdot X_{a_4}+X_{a_1(b_1|}\cdot X_{a_2}\cdot X_{a_3|b_2)}\cdot X_{a_4}\right.\nonumber\\
&&+X_{a_1(b_1|}\cdot X_{a_2}\cdot X_{a_3}\cdot X_{a_4|b_2)}+X_{a_1}\cdot X_{a_2(b_1|}\cdot X_{a_3|b_2)}\cdot X_{a_4}\nonumber\\
&&+\left.X_{a_1}\cdot X_{a_2(b_1|}\cdot X_{a_3}\cdot X_{a_4|b_2)}+X_{a_1}\cdot X_{a_2}\cdot X_{a_3(b_1|}\cdot X_{a_4|b_2)}\right]\,,
\label{56}\\
\partial_{b_1b_2b_3}G_{a_1a_2a_3a_4}&=&\left[X_{a_1b_1b_2b_3}\cdot X_{a_2}\cdot X_{a_3}\cdot X_{a_4}+X_{a_1}\cdot X_{a_2b_1b_2b_3}\cdot X_{a_3}\cdot X_{a_4}\right.\nonumber\\
&&+\left.X_{a_1}\cdot X_{a_2}\cdot X_{a_3b_1b_2b_3}\cdot X_{a_4}+X_{a_1}\cdot X_{a_2}\cdot X_{a_3}\cdot X_{a_4b_1b_2b_3}\right]\nonumber\\
&&+\left[X_{a_1(b_1b_2|}\cdot X_{a_2|b_3)}\cdot X_{a_3}\cdot X_{a_4}+X_{a_1(b_1b_2|}\cdot X_{a_2}\cdot X_{a_3|b_3)}\cdot X_{a_4}\right.\nonumber\\
&&+X_{a_1(b_1b_2|}\cdot X_{a_2}\cdot X_{a_3}\cdot X_{a_4|b_3)}+X_{a_1}\cdot X_{a_2(b_1b_2|}\cdot X_{a_3|b_3)}\cdot X_{a_4}\nonumber\\
&&+X_{a_1}\cdot X_{a_2(b_1b_2|}\cdot X_{a_3}\cdot X_{a_4|b_3)}+X_{a_1}\cdot X_{a_2}\cdot X_{a_3(b_1b_2|}\cdot X_{a_4|b_3)}\nonumber\\
&&+X_{a_1(b_1|}\cdot X_{a_2|b_2b_3)}\cdot X_{a_3}\cdot X_{a_4}+X_{a_1(b_1|}\cdot X_{a_2}\cdot X_{a_3|b_2b_3)}\cdot X_{a_4}\nonumber\\
&&+X_{a_1(b_1|}\cdot X_{a_2}\cdot X_{a_3}\cdot X_{a_4|b_2b_3)}+X_{a_1}\cdot X_{a_2(b_1|}\cdot X_{a_3|b_2b_3)}\cdot X_{a_4}\nonumber\\
&&+\left.X_{a_1}\cdot X_{a_2(b_1|}\cdot X_{a_3}\cdot X_{a_4|b_2b_3)}+X_{a_1}\cdot X_{a_2}\cdot X_{a_3(b_1|}\cdot X_{a_4|b_2b_3)}\right]\nonumber\\
&&+\left[X_{a_1(b_1|}\cdot X_{a_2|b_2|}\cdot X_{a_3|b_3)}\cdot X_{a_4}+X_{a_1(b_1|}\cdot X_{a_2|b_2|}\cdot X_{a_3}\cdot X_{a_4|b_3)}\right.\nonumber\\
&&+\left.X_{a_1(b_1|}\cdot X_{a_2}\cdot X_{a_3|b_2|}\cdot X_{a_4|b_3)}+X_{a_1}\cdot X_{a_2(b_1|}\cdot X_{a_3|b_2|}\cdot X_{a_4|b_3)}\right]\,,
\label{57}\\
\partial_{b_1b_2b_3b_4}G_{a_1a_2a_3a_4}&=&\left[X_{a_1b_1b_2b_3b_4}\cdot X_{a_2}\cdot X_{a_3}\cdot X_{a_4}+X_{a_1}\cdot X_{a_2b_1b_2b_3b_4}\cdot X_{a_3}\cdot X_{a_4}\right.\nonumber\\
&&+\left.X_{a_1}\cdot X_{a_2}\cdot X_{a_3b_1b_2b_3}\cdot X_{a_4}+X_{a_1}\cdot X_{a_2}\cdot X_{a_3}\cdot X_{a_4b_1b_2b_3b_4}\right]\nonumber\\
&&+\left[X_{a_1(b_1b_2b_3|}\cdot X_{a_2|b_4)}\cdot X_{a_3}\cdot X_{a_4}+X_{a_1(b_1b_2b_3|}\cdot X_{a_2}\cdot X_{a_3|b_4)}\cdot X_{a_4}\right.\nonumber\\
&&+X_{a_1(b_1b_2b_3|}\cdot X_{a_2}\cdot X_{a_3}\cdot X_{a_4|b_4)}+X_{a_1}\cdot X_{a_2(b_1b_2b_3|}\cdot X_{a_3|b_4)}\cdot X_{a_4}\nonumber\\
&&+X_{a_1}\cdot X_{a_2(b_1b_2b_3|}\cdot X_{a_3}\cdot X_{a_4|b_4)}+X_{a_1}\cdot X_{a_2}\cdot X_{a_3(b_1b_2b_3|}\cdot X_{a_4|b_4)}\nonumber\\
&&+X_{a_1(b_1|}\cdot X_{a_2|b_2b_3b_4)}\cdot X_{a_3}\cdot X_{a_4}+X_{a_1(b_1|}\cdot X_{a_2}\cdot X_{a_3|b_2b_3b_4)}\cdot X_{a_4}\nonumber\\
&&+X_{a_1(b_1|}\cdot X_{a_2}\cdot X_{a_3}\cdot X_{a_4|b_2b_3b_4)}+X_{a_1}\cdot X_{a_2(b_1|}\cdot X_{a_3|b_2b_3b_4)}\cdot X_{a_4}\nonumber\\
&&+\left.X_{a_1}\cdot X_{a_2(b_1|}\cdot X_{a_3}\cdot X_{a_4|b_2b_3b_4)}+X_{a_1}\cdot X_{a_2}\cdot X_{a_3(b_1|}\cdot X_{a_4|b_2b_3b_4)}\right]\nonumber\\
&&+\left[X_{a_1(b_1b_2|}\cdot X_{a_2|b_3b_4)}\cdot X_{a_3}\cdot X_{a_4}+X_{a_1(b_1b_2|}\cdot X_{a_2}\cdot X_{a_3|b_3b_4)}\cdot X_{a_4}\right.\nonumber\\
&&+X_{a_1(b_1b_2|}\cdot X_{a_2}\cdot X_{a_3}\cdot X_{a_4|b_3b_4)}+X_{a_1}\cdot X_{a_2(b_1b_2|}\cdot X_{a_3|b_3b_4)}\cdot X_{a_4}\nonumber\\
&&+\left.X_{a_1}\cdot X_{a_2(b_1b_2|}\cdot X_{a_3}\cdot X_{a_4|b_3b_4)}+X_{a_1}\cdot X_{a_2}\cdot X_{a_3(b_1b_2|}\cdot X_{a_4|b_3b_4)}\right]\nonumber\\
&&+\left[X_{a_1(b_1b_2|}\cdot X_{a_2|b_3|}\cdot X_{a_3|b_4)}\cdot X_{a_4}+X_{a_1(b_1b_2|}\cdot X_{a_2|b_3|}\cdot X_{a_3}\cdot X_{a_4|b_4)}\right.\nonumber\\
&&+X_{a_1(b_1b_2|}\cdot X_{a_2}\cdot X_{a_3|b_3|}\cdot X_{a_4|b_4)}+X_{a_1(b_1|}\cdot X_{a_2|b_2b_3|}\cdot X_{a_3|b_4)}\cdot X_{a_4}\nonumber\\
&&+X_{a_1(b_1|}\cdot X_{a_2|b_2b_3|}\cdot X_{a_3}\cdot X_{a_4|b_4)}+X_{a_1}\cdot X_{a_2(b_1b_2|}\cdot X_{a_3|b_3|}\cdot X_{a_4|b_4)}\nonumber\\
&&+X_{a_1(b_1|}\cdot X_{a_2|b_2|}\cdot X_{a_3|b_3b_4)}\cdot X_{a_4}+X_{a_1(b_1|}\cdot X_{a_2}\cdot X_{a_3|b_2b_3|}\cdot X_{a_4|b_4)}\nonumber\\
&&+X_{a_1}\cdot X_{a_2(b_1b_2|}\cdot X_{a_3|b_3|}\cdot X_{a_4|b_4)}+X_{a_1(b_1|}\cdot X_{a_2|b_2|}\cdot X_{a_3}\cdot X_{a_4|b_3b_4)}\nonumber\\
&&+\left.X_{a_1(b_1|}\cdot X_{a_2}\cdot X_{a_3|b_2|}\cdot X_{a_4|b_3b_4)}+X_{a_1}\cdot X_{a_2(b_1|}\cdot X_{a_3|b_2|}\cdot X_{a_4|b_3b_4)}\right]\nonumber\\
&&+X_{a_1(b_1|}\cdot X_{a_2|b_2|}\cdot X_{a_3|b_3|}\cdot X_{a_4|b_4)}\,.
\label{58}
\end{eqnarray}

\noindent Then, the derivatives above contain terms of the form

\begin{eqnarray}
\partial^4{\bf G}&=&(\partial^4\Lambda)\,\Lambda^3+(\partial^3\Lambda)\,(\partial\Lambda)\,\Lambda^2
+(\partial^2\Lambda)^2\,\Lambda^2+(\partial^2\Lambda)\,(\partial\Lambda)^2\,\Lambda+(\partial\Lambda)^4\,,
\label{591}\\
\partial^3{\bf G}&=&(\partial^3\Lambda)\,\Lambda^3+(\partial^2\Lambda)\,(\partial\Lambda)\,\Lambda^2+(\partial\Lambda)^3\,\Lambda\,,
\label{592}\\
\partial^2{\bf G}&=&(\partial^2\Lambda)\,\Lambda^3+(\partial\Lambda)^2\,\Lambda^2\,,
\label{593}\\
\partial{\bf G}&=&(\partial\Lambda)\,\Lambda^3\,.
\label{594}
\end{eqnarray}

\noindent where $\Lambda=\partial X$.

\bigskip

The way in which the different terms are combined is given by the comments following (\ref{20}). Then, we have

\begin{equation}
R_{i_1i_2i_3i_4j_1j_2j_3j_4}=S_{i_1i_2i_3i_4j_1j_2j_3j_4}+({\rm something})\,,
\label{60}
\end{equation}

\noindent where

\begin{eqnarray}
S_{i_1i_2i_3i_4j_1j_2j_3j_4}&=&\partial_{i_1i_2i_3i_4}G_{j_1j_2j_3j_4}\nonumber\\
&&-\left(\partial_{j_1i_2i_3i_4}G_{i_1j_2j_3j_4}+\partial_{i_1j_2i_3i_4}G_{j_1i_2j_3j_4}\right.\nonumber\\
&&\quad+\left.\partial_{i_1i_2j_3i_4}G_{j_1j_2i_3j_4}+\partial_{i_1i_2i_3j_4}G_{j_1j_2j_3i_4}\right)\nonumber\\
&&+\left(\partial_{j_1j_2i_3i_4}G_{i_1i_2j_3j_4}+\partial_{j_1i_2j_3i_4}G_{i_1j_2i_3j_4}
+\partial_{j_1i_2i_3j_4}G_{i_1j_2j_3i_4}\right.\nonumber\\
&&\quad+\left.\partial_{i_1j_2j_3i_4}G_{j_1i_2i_3j_4}+\partial_{i_1j_2i_3j_4}G_{j_1i_2j_3i_4}
+\partial_{i_1i_2j_3j_4}G_{j_1j_2i_3i_4}\right)\nonumber\\
&&-\left(\partial_{i_1j_2j_3j_4}G_{j_1i_2i_3i_4}+\partial_{j_1i_2j_3j_4}G_{i_1j_2i_3i_4}\right.\nonumber\\
&&\quad+\left.\partial_{j_1j_2i_3j_4}G_{i_1i_2j_3i_4}+\partial_{j_1j_2j_3i_4}G_{i_1i_2i_3j_4}\right)\nonumber\\
&&+\partial_{j_1j_2j_3j_4}G_{i_1i_2i_3i_4}\,.
\label{59}
\end{eqnarray}

\noindent and ``something'' contains derivatives of ${\bf G}$ of lesser order in a combination such as to cancel all transformation matrices appearing there.

It must be by now clear the kind of algebraic manipulations necessary to construct the desired invariant and that they are quite involved. Work is in progress to develop a calculational algorithm to determine the full expression of the differential invariant involving the non--linear terms and the inverse tensor (\ref{49}).

\section{Conclusions}

We have outlined the construction of differential invariants for higher--rank tensors.

\section*{Acknowledgements}

This work was partially done at the {\bf Abdus Salam} International Centre for Theoretical Physics, Trieste. The work was partially supported by Direcci\'on de Investigaci\'on --- Bogot\'a, Universidad Nacional de Colombia.



\begin{thebibliography}{XX}

\bibitem{ABH}M. Atiyah, R. Bott and V. K. Hodge, {\it On the heat equation and the index theorem}, Inventiones Math. {\bf 19}, 279 ({\tt 1973}).

\bibitem{BB}A. K. H. Bengtsson and I. Bengtsson, {\it Massless higher--spin fields revisited}, Class. Quantum Grav. {\bf 3}, 927 ({\tt 1986}).

\bibitem{BBV}F. A. Berends, G. J. H. Burgers and H. van Dam, {\it On spin three self interactions}, Z. Phys. C {\bf 24}, 247 ({\tt 1984}).

\bibitem{Cur}T. Curtright, {\it Generalized gauge fields}, Phys. Lett. B {\bf 165}, 304 ({\tt 1985}).

\bibitem{DD1}T. Damour and S. Deser, {\it Higher--derivative interactions of higher--spin gauge fields}, Class. Quantum Grav. {\bf 4}, L95 ({\tt 1987}).

\bibitem{DD2}T. Damour and S. Deser, {\it ``Geometry'' of spin 3 gauge theories}, Ann. Inst. Henri Poincar\'e {\bf 47}, 277 ({\tt 1987}).

\bibitem{DF}B. de Wit and D. Z. Freedman, {\it Systematics of higher--spin gauge fields}, Phys. Rev. D {\bf 21}, 358 ({\tt 1980}).

\bibitem{FF}J. Fang and C. Fronsdal, {\it Massless fields with half--integer spin}, Phys. Rev. D {\bf 18}, 3630 ({\tt 1978}).

\bibitem{FS}D. Francia and A. Sagnotti, {\it Free geometric equations for higher spins}, Phys. Lett. B {\bf 543}, 303 ({\tt 2002}).

\bibitem{Fro}C. Fronsdal, {\it Massless fields with integer spin}, Phys. Rev. D {\bf 18}, 3624 ({\tt 1978}).

\bibitem{FKWC}S. A. Fulling, R. C. King, B. G. Wybourne and C. J. Cummins, {\it Normal forms for tensor polynomials: I. The Riemann tensor}, Class. Quantum Grav. {\bf 9}, 1151 ({\tt 1992}).

\bibitem{Gau}K. F. Gauss, {\it Disquisitiones generales circa superficies curvas} ({\tt 1827}).

\bibitem{Hul}C. M. Hull, {\it Duality and gravity in higher spin gauge fields}, J. High Energy Phys. {\bf 09}, 027 ({\tt 2001}).

\bibitem{MH1}P. F. Medeiros and C. M. Hull, {\it Exotic tensor gauge theory and duality}, Commun. Math. Phys. {\bf 235}, 255 ({\tt 2003}).

\bibitem{MH2}P. Medeiros and C. M. Hull, {\it Geometric second order field equations for general tensor gauge fields}, J. High Energy Phys. {\bf 05}, 019 ({\tt 2003}).

\bibitem{MV}J. Mu\~noz Masqu\'e and A. Vald\'es, {\it The number of functionally independent invariants of a pseudo--Riemannian metric}, J. Phys. A: Math. Gen. {\bf 27}, 7843 ({\tt 1994}).

\bibitem{Rie}B. Riemann, {\it Commentatio mathematica, qua respondere tentatur quaestioni ab Ill$^{ma}$ Academia Parisiensi propositae} ({\tt 1861}); Reprinted in {\it Gesammelte matematische Werke} (Teubner, Lei\-pzig, {\tt 1892}; Dover, New York, {\tt 1953}); English traslation in R. Farwell and C. Knee, {\it The missing link: Riemann's ``Commentatio,'' differential geometry and tensor analysis}, Hist. Math. {\bf 17}, 223 ({\tt 1990}).

\bibitem{Ta1}V. Tapia, {\it Integrable conformal field theory in four dimensions and fourth--rank geometry}, Int. J. Mod. Phys. D {\bf 3}, 413 ({\tt 1993}).

\bibitem{TRMC}V. Tapia, D. K. Ross, A. L. Marrakchi and M. Cataldo, {\it Renormalizable conformally invariant model for the gravitational field}, Class. Quantum Grav. {\bf 13}, 3261 ({\tt 1996}).

\bibitem{TR}V. Tapia and D. K. Ross, {\it Conformal fourth--rank gravity, non--vanishing cosmological constant, and anisotropy}, Class. Quantum Grav. {\bf 15}, 245 ({\tt 1998}).

\bibitem{Ta3}V. Tapia, {\it Algebraic invariants, determinants, and Cayley--Hamilton theorem for hypermatrices: the fourth--rank case}, math--ph/0208010 ({\tt 2002}).

\bibitem{Tho}T. Y. Thomas, {\it The Differential Invariants of Generalized Spaces} (Cambridge University Press, Cambridge, {\tt 1934}).

\bibitem{Vas}M. A. Vasiliev, {\it ``Gauge'' form of descrption of masless fields with arbitrary spin}, Sov. J. Nucl. Phys. {\bf 32}, 439 ({\tt 1980}).

\bibitem{WM}B. G. Wybourne and J. Meller, {\it Enumeration of the order--14 invariants formed from the Riemann tensor}, J. Phys. A: Math. Gen. {\bf 25}, 5999 ({\tt 1992}).

\end{thebibliography}
\end{document}